\documentclass[letterpaper, 10 pt, conference]{ieeeconf}

\IEEEoverridecommandlockouts

\usepackage{cite}
\usepackage{url}
\usepackage{amsmath,amssymb,amsfonts}
\usepackage{algorithmic}
\usepackage{graphicx}
\usepackage{adjustbox}
\usepackage{textcomp}
\usepackage{multirow}
\usepackage{comment}
\usepackage[table]{xcolor}
\usepackage{lipsum}
\usepackage{subfiles}
\usepackage{subcaption}
\usepackage{booktabs}

\definecolor{rowgray}{gray}{0.9}
\def\BibTeX{{\rm B\kern-.05em{\sc i\kern-.025em b}\kern-.08em
    T\kern-.1667em\lower.7ex\hbox{E}\kern-.125emX}}

\DeclareMathOperator*{\argmax}{arg\,max}

\begin{document}

\title{\LARGE \bf Deep Reinforcement Learning for Misbehavior Detection \\Under Partially Observable V2X Data
}

\author{Roshan Sedar and Charalampos Kalalas 
\thanks{This work has been partially supported by the project PID2024-160944OB-I00 (SEASIDE) funded by MICIU/AEI/10.13039/501100011033 and the internal project BARYCENTRE (300960).}
\thanks{Roshan Sedar and Charalampos Kalalas are with the Sustainable Artificial Intelligence Research Unit,
        Centre Tecnològic de Telecomunicacions de Catalunya (CTTC/CERCA), 08860 Castelldefels, Spain
        {\tt\small  \{roshan.sedar, ckalalas\}@cttc.es}%
}
}

\maketitle
\thispagestyle{empty}
\pagestyle{empty}

\begin{abstract}
Misbehavior detection in vehicle-to-everything (V2X) systems is essential for ensuring the semantic correctness of exchanged messages and preventing the dissemination of falsified information.
Existing data-centric misbehavior detection approaches largely rely on statistical validation or supervised machine learning models under the implicit assumption of fully observable V2X streams.
In practice, however, vehicular environments are inherently partially observable due to hardware failures, intermittent connectivity, and environmental occlusions. Moreover, missingness itself can be strategically exploited by adversaries to evade detection.
In this paper, we study misbehavior detection under incomplete V2X observations and propose a deep reinforcement learning (DRL)-based detection framework that learns adaptive policies with incomplete data.
We further introduce an adversarial threat model in which attackers exploit or deliberately induce missingness to evade detection, including evasion via natural occlusions and adversarial feature suppression.
Extensive experiments conducted on the VeReMi dataset under various missingness patterns demonstrate that DRL significantly outperforms a powerful XGBoost baseline under natural partial observability. However, results also reveal a critical vulnerability: DRL policies can be highly susceptible to evasion attacks that strategically exploit natural missingness. In contrast, DRL exhibits more gradual degradation under direct feature suppression compared to static tree-based models.

\end{abstract}

\vspace{3pt}

\section{Introduction}
Misbehavior detection has recently become one of the primary concerns in emerging vehicle-to-everything (V2X) systems characterized by pervasive sensing, computing, and connectivity capabilities. Traditional cryptographic mechanisms, while essential for authentication and integrity, are insufficient to guarantee the semantic correctness of transmitted data. As a result, data-centric misbehavior detection relies on the analysis of spatiotemporal measurement streams, captured at different locations and time instances, to 
determine the trustworthiness of information by detecting abnormal behaviors with the aid of statistical analysis, rule-based validation, or machine learning (ML) models \cite{10015746}.

Nevertheless, a fundamental yet often overlooked challenge for 
reliable misbehavior detection resides in the completeness of aggregated information. 
In practice, the emergence of missing/incomplete data in the fused vehicular measurement streams is inevitable \cite{10.1109/TITS.2024.3478816}. Partial observability of mobility information can be generally attributed to the following factors: i) \textit{hardware failures}, where the malfunction of vehicle components (e.g., synchronization failures or errors in sensor readings) may result in persistent missing observations for one or multiple state variables of the vehicle; ii) \textit{intermittent connectivity}, as a result of the high mobility, dynamic network topology, and wireless channel impairments 
which may result in sporadic outages and packet losses for consecutive time-steps; 
and iii) \textit{environmental factors}, such as buildings, large vehicles, adverse weather conditions, and road curvature, which often introduce occlusions, leading to incomplete situational awareness. Collectively, these factors result in a partially observable V2X environment in which detection methods must reason under significant epistemic uncertainty about the true system state.

The challenge of partial observability becomes particularly relevant when malicious actors exploit missingness to intentionally obscure inconsistencies (e.g., via stealthy data injections or subtle perturbations) and evade detection. In such scenarios, the adversarial exploitation of missingness aims to degrade the effectiveness of misbehavior detection systems and compromise their trustworthiness. 
Existing research on data-centric misbehavior detection includes plausibility-based detectors \cite{tsukada2022misbeh}, consistency checks across neighboring vehicles \cite{9062831}, trust and reputation schemes \cite{9099055}, as well as supervised \cite{prinkle21MLMBD, 9662982} and unsupervised \cite{Campos2025} ML models. Despite this growing body of work, the majority of these methods implicitly assume either fully observable V2X streams or the straightforward exclusion of unobserved samples to overcome data incompleteness.
Thus, the effectiveness of misbehavior detection under partial observability remains unexplored.

In \cite{9608954}, the authors aim to answer whether imputation or missing-tolerant classification yields the best misbehavior detection performance in incomplete V2X streams. Missingness is synthetically induced with different ratios, mechanisms, and distributions using a multi-factor amputation framework that allows a comprehensive benchmark comparison of missing data handling strategies. However, missing data is solely treated as a static classification challenge with merely incomplete feature vectors and without ensuring sequential reasoning over time, since learning relies on supervised classification. As such, adaptive detection decisions (e.g., policy learning instead of fixed classification outputs) cannot be guaranteed. 
Additionally, to the best of our knowledge, no prior work 
considers \textit{adversarially induced missingness} as a deliberate attack vector against misbehavior detection models. 
This gap leaves deployed detection systems fundamentally exposed to a class of evasion threats for which they were not designed.
All these factors underscore the need for a dedicated study that explicitly accounts for partially observed V2X environments caused by both natural occlusions and adversarial manipulation.

To bridge these gaps, this paper proposes a novel deep reinforcement learning (DRL) framework for misbehavior detection specifically designed for partially observable vehicular environments. 
Unlike prior work that treats missingness as a preprocessing task or static classification problem,
we reframe misbehavior detection as a sequential decision-making problem that allows reasoning under incomplete vehicular streams.  
Detection robustness is empirically evaluated using the VeReMi benchmark dataset \cite{kamel2020veremi} by considering three scenarios: \textit{i}) natural occlusions via standard missingness patterns; \textit{ii}) adversarial evasion that exploits naturally occurring missing regions to inject stealthy perturbations; and \textit{iii})
adversarial feature suppression that actively induces missingness as an attack strategy.
A comprehensive benchmark assessment against an XGBoost baseline reveals both the robustness advantages and security vulnerabilities of DRL under partial observability.

The remainder of this paper is organized as follows. Section \ref{sec:two} introduces the considered missingness scenarios in our study. Section \ref{sec:misdet} presents the proposed DRL-based misbehavior detection mechanism and the adversarial threat model. Section \ref{sec:exp} details the experimental setup, while Section \ref{sec:res} provides a comprehensive performance evaluation and benchmark comparison of the proposed framework across all considered scenarios. Finally, Section~\ref{sec:concl} concludes the paper and lists future work.

\section{Missingness Scenarios} \label{sec:two}
\begin{figure}[!t]
    \centering
    \begin{subfigure}[b]{0.51\textwidth}
        \centering
        \includegraphics[width=\textwidth, keepaspectratio]{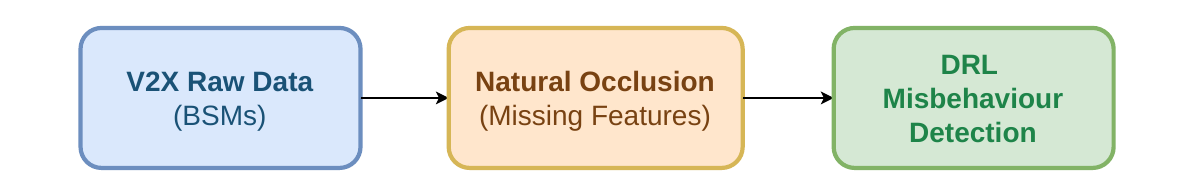}
        \caption{Natural occlusions.}
        \label{fig:uc1}
    \end{subfigure}
    \hfill
    \begin{subfigure}[b]{0.49\textwidth}
        \centering
        \includegraphics[width=\textwidth, keepaspectratio]{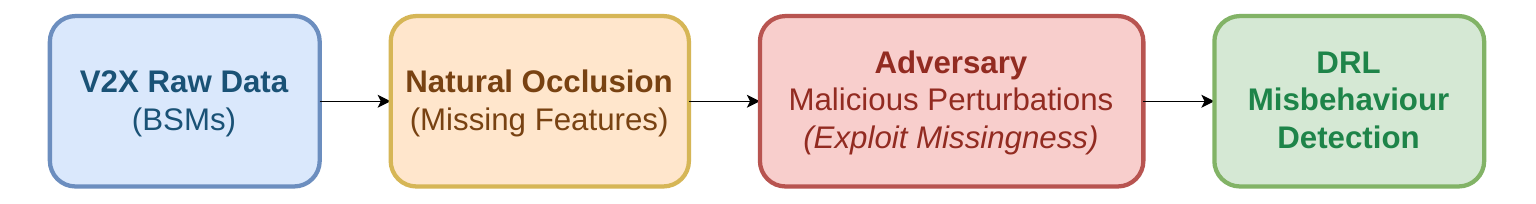}
        \caption{Natural occlusions and adversarial perturbations.}
        \label{fig:uc2}
    \end{subfigure}
    \hfill
    \begin{subfigure}[b]{0.5\textwidth}
        \centering
        \includegraphics[width=\textwidth, keepaspectratio]{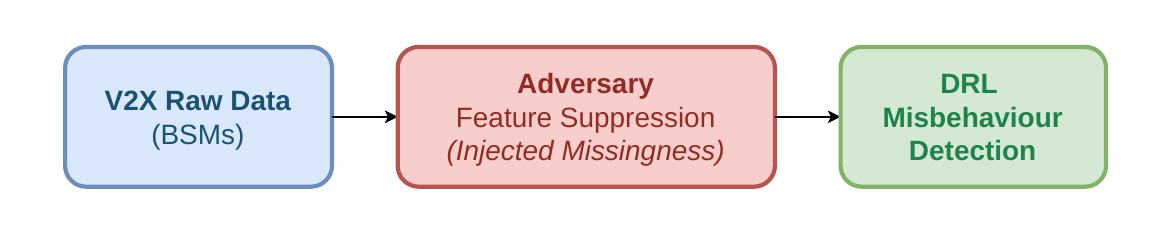}
        \caption{Adversarial feature suppression.}
        \label{fig:uc3}
    \end{subfigure}

    \caption{Considered scenarios under data missingness.} 
    \label{fig:usecases}
    \vspace{-0.45cm}
\end{figure}

We consider a set of missingness scenarios, as illustrated in Fig.~\ref{fig:usecases}, to systematically evaluate the misbehavior detection capabilities of the proposed DRL-based framework under partial observability. 
The considered scenarios progressively transition from natural missingness to adversarially induced suppression, thereby enabling a comprehensive assessment of detection performance
across diverse and challenging observability conditions.
Next, we elaborate on the considered scenarios.

\textbf{SC1: Natural occlusions.} Due to the dynamic and dense nature of vehicular environments, data missingness in V2X streams is inherent and unavoidable. Complete and synchronized perception of the vehicular system state is thus rarely attainable in practice. As such, the first scenario, shown in Fig.~\ref{fig:uc1}, models partial observability arising from non-malicious hardware, communication, and environmental factors. In real-world vehicular networks, natural occlusions stem from transient sensor failures, multipath fading, physical obstruction between vehicles or buildings, packet collisions, and adverse weather conditions.

\textbf{SC2: Evasion via natural occlusions.} Building on SC1, this scenario aims to demonstrate that partial observability not only poses a technical challenge for misbehavior detection systems but simultaneously enlarges the attack surface available to adversarial actors. In particular, SC2 considers an exogenous adversary that exploits the naturally occurring missingness in V2X data streams to launch evasion attacks against the DRL framework. To realize the attack, the adversary leverages existing missing regions induced by natural occlusions, rather than introducing new missing features. 
By timing subtle false data injections to coincide with periods of natural occlusion or by crafting adversarial perturbations calibrated to blend with the statistical profile of legitimate blind spots,
as shown in Fig.~\ref{fig:uc2}, the adversary seeks to minimize its detectability while maximizing the likelihood that the detector misclassifies misbehaving vehicles as benign.

\textbf{SC3: Adversarial feature suppression.} This scenario considers an exogenous adversary that actively induces missingness as an attack strategy rather than exploiting existing missing regions. Thus, in contrast to SC1 and SC2, this scenario represents adversarially induced partial observability where missingness itself becomes the attack vector.
The original V2X data streams in SC3 are assumed to be complete, and natural occlusions have been addressed through appropriate imputation techniques, prior to adversarial intervention.
As shown in Fig.~\ref{fig:uc3}, the adversary executes a feature suppression attack by selectively injecting artificially induced missing values across input features of the V2X data stream. The goal is to reduce the information available to the detector and degrade the classifier's ability to distinguish between normal and misbehaving vehicles.

\section{Misbehavior Detection} 
\label{sec:misdet}

\subsection{Model}

The V2X environment considered in this study follows a Markov decision process (MDP) framework to facilitate the detection of misbehaviors through sequential decision-making. Consistent with the MDP formulation, the action of misbehavior detection changes the environment based on the decision of either legitimate or malicious behavior at time-step $t$. Subsequently, the decision at time-step $t+1$ is influenced by the altered environment from the previous time-step $t$.
%
%
%
In this work, we consider a DRL-based detector deployed at an edge node. 
%

\textit{i) Agent:} The agent takes the V2X data as a time series and prior decisions as its state $\mathbf{s_{t}}$, and outputs an action $a_{t}$ according to policy $\pi$. The agent's DQN~\cite{mnih2015human} consists of an LSTM layer followed by a fully connected network with linear activation to estimate the $Q$-values over possible actions. Each interaction is stored as a transition tuple $ e_{t} = <\mathbf{s_{t}},a_{t},r_{t},\mathbf{s_{t+1}}>$, capturing the full behavioral trace of the misbehavior detector. By replaying this experience, the agent progressively refines its $Q(\mathbf{s},a)$ estimates towards more accurate misbehavior detection. The objective is to maximize the expected sum of future discounted rewards, expressed as 
$R_{t} = \sideset{}{}\sum_{k=t}^{T} \gamma^{k-t}r_{k}.$
The $Q$-values are updated at each step using learning rate $\alpha$ and discount factor $\gamma$ as
\begin{align}
\begin{split}
Q(\mathbf{s_{t}}, a_{t}) &\leftarrow Q(\mathbf{s_{t}}, a_{t})\\
 &+\alpha(r_{t} + \gamma\max_{a_{t+1}} Q(\mathbf{s_{t+1}}, a_{t+1}) - Q(\mathbf{s_{t}}, a_{t})).
\end{split}
\end{align}%

\textit{ii) States:} The state $\mathbf{s_{t}}$ comprises two components: a sequence of prior actions $s_{action} = <a_{t-1},a_{t},...,a_{t+n-1}>$ and a window of current BSM observations $s_{time} = <\mathbf{X_{t}},...,\mathbf{X_{t+n}}>$, where $\mathbf{X_{t}} \in \mathbb{R}^d$ is a $d$-dimensional feature vector at time-step $t$. This design allows the agent to capture temporal dependencies across both past decisions and incoming BSM data, enabling more informed action selection. 

\textit{iii) Actions:} The action space is $\mathcal{A}$ = \{0,1\}, where $1$ indicates detected misbehavior and $0$ represents the genuine behavior. The deterministic policy $\pi:\mathbf{s_{t}} \in \mathcal{S} \longmapsto a_{t} \in \mathcal{A}$ maps each state to a single action.  
In state $\mathbf{s_{t}}$, the agent selects the action that maximizes the optimal $Q$-function, 
\begin{align}
        \pi^* = \argmax\limits_{a \in \mathcal{A}} Q^*(\mathbf{s},a).
        \label{eq:eps_greedy}
\end{align}

\textit{iv) Rewards:}  
The reward $r_t$ is defined over the four outcomes of the confusion matrix typically used in ML classification problems. Correct decisions yield positive rewards, while errors are penalized, with false negatives (FNs) penalized more heavily than false positives (FPs), since missed misbehavior poses a higher safety risk. Accordingly, the reward of the agent is 
\begin{align}
r(\mathbf{s_{t}},a_{t}) = \begin{cases}
                a, & \text{if}~a_{t}~\text{is a TP,}\\
                b, & \text{if}~a_{t}~\text{is a TN,}\\
                -c, & \text{if}~a_{t}~\text{is an FP,}\\
                -d, & \text{if}~a_{t}~\text{is an FN,}
            \end{cases}
\label{eq:rew_func}
\end{align}
where $a, b, c, d > 0$,  with $a > b$ and $d > c$. 

\subsection{Adversarial Threat Model}

We assume the presence of black-box adversaries attempting to evade the DRL-based misbehavior detection system through malicious manipulation of V2X data, by exploiting or introducing missing features. The adversary is assumed to have sufficient knowledge of the data pipeline and the first-order statistical properties of the input data distribution, such as mean values and standard deviations of the feature vector, which capture the central tendencies and dispersion characteristics of legitimate V2X data~\cite{papernot2016limitations,ilyas2019adversarial}. The attacker does not require direct access to the internal parameters of the DRL model. In this context, we consider an exogenous attacker capable of executing two adversarial strategies, each representing a distinct threat vector targeting the integrity of the input data. Next, the two adversarial strategies constituting the adversarial model are described.

\textbf{Evasion via natural occlusions.} In this case, the adversary leverages naturally occurring missing regions within the V2X data stream
to inject adversarial perturbations that are crafted to blend with the underlying occluded data distribution (shown in Fig.~\ref{fig:uc2}). We assume that the adversary strategically targets high-variance features to maximize perturbation stealthiness, as these regions exhibit high natural variability that masks manipulation~\cite{papernot2016limitations,ilyas2019adversarial}. The adversarial input vector $\boldsymbol{\epsilon}_{\text{adv}} = C + \boldsymbol{\rho}$ is derived from the benign mean vector $C = \boldsymbol{\mu}$ of the training distribution, where perturbations $\rho_i = \text{sign}(C_i) \cdot k \cdot \sigma_i$ are applied to the top-$n = \lfloor d \cdot r \rfloor$ high-variance features ($d$: feature dimension, $r$: missingness rate, $k$: perturbation scale). The resulting adversarial values $\boldsymbol{\epsilon}_{\text{adv}}$ then fill the missing positions. By doing this, the adversarial examples become indistinguishable from legitimate incomplete observations, penetrating the DRL classifier undetected. This forms an evasion attack where the adversary exploits the inherent issues of the V2X environment as a camouflage for adversarial manipulation.    

\textbf{Adversarial feature suppression.} In this case, the adversary actively introduces missingness as an attack strategy rather than exploiting naturally occurring occlusions. The original V2X data streams are complete with no missing feature vectors, as in both SC1 and SC2 before amputation. As shown in Fig.~\ref{fig:uc3}, the adversary executes a feature suppression attack by selectively injecting missing values or blocking a subset of input features within the V2X data stream, intending to achieve controlled information suppression.
Specifically, missingness is injected at a rate proportional to $\lambda \in [0, 1]$, ranging from 1\% to 20\%, remaining within plausible natural missingness regions similar to the patterns evaluated in SC1 while maintaining feature suppression stealthiness. The resulting missing entries are then filled with adversarial inputs $x^{adv}_i = x_i + \epsilon_i$, where $\epsilon_i = (\mu_i - \delta_{\lambda} \cdot \sigma_i) - x_i$ drives the feature toward a value between $1.8\sigma_i$ and $3.5\sigma_i$ below $\mu_i$, a range empirically derived to maximize classifier degradation while maintaining statistical plausibility, with an additive Gaussian noise to reduce detectability~\cite{papernot2016limitations}. This strategy artificially suppresses the informational content of targeted features, degrading the DRL classifier's ability to distinguish between normal and misbehaving vehicles.

\section{Experiments} \label{sec:exp}

\subsection{VeReMi Dataset}
In this study, we utilize the open-source VeReMi dataset~\cite{kamel2020veremi} to assess misbehavior detection performance. The dataset provides labeled BSMs spanning multiple misbehavior types and traffic densities, making it a well-established benchmark for V2X security research. A representative subset of misbehaviors is selected to provide sufficient coverage of the available misbehavior types in VeReMi, as summarized below. 

\textbf{Position falsification attack:} A vehicle transmits falsified position coordinates within its communication area while concealing its true position. Among the possible variants, three are considered here: i) \textit{constant position}, where a fixed position coordinate is repeatedly broadcast; ii) \textit{constant position offset}, where the true position is transmitted with a fixed offset; and iii) \textit{random position}, where newly generated random coordinates are broadcast at each transmission.

\textbf{Speed falsification attack:} A misbehaving vehicle transmits falsified speed values in its BSMs, following a similar approach to position falsification. Among the possible variants, one is considered here: \textit{random speed}, where newly generated random speed values are broadcast at each transmission.

\textbf{Delayed messages:} A misbehaving vehicle transmits BSMs with correct field values but introduces a deliberate time delay, shifting transmissions away from real-time reporting.

\begin{table*}[!ht]
\centering
\caption{Detection performance for DRL and XGBoost under natural occlusions in SC1.}
\label{tab:detection_perf_sc1}
\begin{tabular}{l c |cc cc cc cc| cc cc cc cc cc}
    \toprule
    \rowcolor{rowgray}
    & & \multicolumn{8}{c}{\textbf{DRL}} & \multicolumn{8}{c}{\textbf{XGBoost}} \\\cline{3-10} \cline{11-18}
    \rowcolor{rowgray}
    \textbf{Misbehavior} & \textbf{Missingness} 
    & \multicolumn{2}{c}{\textbf{Normal}}
    & \multicolumn{2}{c}{\textbf{MCAR}} 
    & \multicolumn{2}{c}{\textbf{MNAR}} 
    & \multicolumn{2}{c|}{\textbf{MAR}} 
    & \multicolumn{2}{c}{\textbf{Normal}} 
    & \multicolumn{2}{c}{\textbf{MCAR}} 
    & \multicolumn{2}{c}{\textbf{MNAR}} 
    & \multicolumn{2}{c}{\textbf{MAR}} 
    \\
    \rowcolor{rowgray}
    & 
    & \textbf{Acc} & \textbf{F1} 
    & \textbf{Acc} & \textbf{F1} 
    & \textbf{Acc} & \textbf{F1} 
    & \textbf{Acc} & \textbf{F1} 
    & \textbf{Acc} & \textbf{F1} 
    & \textbf{Acc} & \textbf{F1} 
    & \textbf{Acc} & \textbf{F1} 
    & \textbf{Acc} & \textbf{F1} \\
    \midrule
    \multirow{4}{*}{\begin{tabular}[l]{@{}c@{}}Constant \\ Position\end{tabular}}
    & {0\%}
    & 0.99 & 0.98 & $-$ & $-$ & $-$ & $-$ & $-$ & $-$ & 0.92 & 0.88 & $-$ & $-$ & $-$ & $-$ & $-$ & $-$ \\

    & {1\%}
    & $-$ & $-$ & 0.98 & 0.96 & 0.98 & 0.97 & 0.98 & 0.97 & $-$ & $-$ & 0.91 & 0.85 & 0.92 & 0.87 & 0.92 & 0.87   \\

    & {5\%}
    & $-$ & $-$ & 0.92 & 0.85 & 0.97 & 0.96 & 0.98 & 0.97 & $-$ & $-$ & 0.83 & 0.76 & 0.90 & 0.84 & 0.91 & 0.85   \\

    & {10\%}
    & $-$ & $-$ & 0.86 & 0.70 & 0.97 & 0.95 & 0.98 & 0.96 & $-$ & $-$ & 0.75 & 0.69 & 0.89 & 0.84 & 0.89 & 0.83   \\

    & {20\%}
    & $-$ & $-$ & 0.82 & 0.58 & 0.95 & 0.91 & 0.96 & 0.93 & $-$ & $-$ & 0.62 & 0.59 & 0.88 & 0.81 & 0.85 & 0.79   \\
 
    \midrule
    \multirow{4}{*}{\begin{tabular}[l]{@{}c@{}}Constant \\ Position \\Offset\end{tabular}}
    & {0\%}
    & 0.99 & 0.98 & $-$ & $-$ & $-$ & $-$ & $-$ & $-$ & 0.78 & 0.72 & $-$ & $-$ & $-$ & $-$ & $-$ & $-$ \\

    & {1\%}
    & $-$ & $-$ & 0.98 & 0.97 & 0.99 & 0.98 & 0.98 & 0.97 & $-$ & $-$ & 0.78 & 0.72 & 0.78 & 0.73 & 0.78 & 0.73   \\

    & {5\%}
    & $-$ & $-$ & 0.95 & 0.91 & 0.98 & 0.96 & 0.98 & 0.96 & $-$ & $-$ & 0.77 & 0.70 & 0.78 & 0.72 & 0.78 & 0.73   \\

    & {10\%}
    & $-$ & $-$ & 0.93 & 0.88 & 0.97 & 0.95 & 0.96 & 0.94 & $-$ & $-$ & 0.76 & 0.67 & 0.78 & 0.72 & 0.78 & 0.72   \\

    & {20\%}
    & $-$ & $-$ & 0.89 & 0.81 & 0.96 & 0.93 & 0.94 & 0.91 & $-$ & $-$ & 0.73 & 0.60 & 0.77 & 0.70 & 0.77 & 0.72   \\

    \midrule
    \multirow{4}{*}{\begin{tabular}[l]{@{}c@{}}Random \\ Position\end{tabular}}
    & {0\%}
    & 0.98 & 0.97 & $-$ & $-$ & $-$ & $-$ & $-$ & $-$ & 0.83 & 0.78 &  $-$ & $-$ & $-$ & $-$ & $-$ & $-$ \\

    & {1\%}
    & $-$ & $-$ & 0.95 & 0.92 & 0.96 & 0.92 & 0.96 & 0.92 & $-$ & $-$ & 0.82 & 0.76 & 0.83 & 0.77 & 0.84 & 0.78 \\

    & {5\%}
    & $-$ & $-$ & 0.94 & 0.88 & 0.95 & 0.92 & 0.96 & 0.92 & $-$ & $-$ & 0.77 & 0.72 & 0.82 & 0.76 & 0.83 & 0.77 \\

    & {10\%}
    & $-$ & $-$ & 0.92 & 0.85 & 0.95  & 0.91 & 0.96 & 0.93 & $-$ & $-$ & 0.72 & 0.67 & 0.80 & 0.76 & 0.82 & 0.76 \\

    & {20\%}
    & $-$ & $-$ & 0.92 & 0.86 & 0.93 & 0.88 & 0.95 & 0.92 & $-$ & $-$ & 0.62 & 0.60 & 0.80 & 0.74 & 0.81 & 0.75 \\
    \midrule
    \multirow{4}{*}{\begin{tabular}[l]{@{}c@{}}Random \\ Speed\end{tabular}}
   & {0\%}
    & 1.0 & 1.0 & $-$ & $-$ & $-$ & $-$ & $-$ & $-$ & 0.99 & 0.98 & $-$ & $-$ & $-$ & $-$ & $-$ & $-$ \\

    & {1\%}
    & $-$ & $-$ & 0.97 & 0.95 & 0.98 & 0.96 & 0.97 & 0.95 & $-$ & $-$ & 0.98 & 0.97 & 0.98 & 0.98 & 0.98 & 0.98 \\

    & {5\%}
    & $-$ & $-$ & 0.95 & 0.91 & 0.96 & 0.93 & 0.96 & 0.93 & $-$ & $-$ & 0.95 & 0.93 & 0.98 & 0.96 & 0.98 & 0.98 \\

    & {10\%}
    & $-$ & $-$ & 0.93 & 0.86 & 0.96 & 0.92 & 0.94 & 0.89 & $-$ & $-$ & 0.92 & 0.88 & 0.97 & 0.95 & 0.98 & 0.97 \\

    & {20\%}
    & $-$ & $-$ & 0.89 & 0.77 & 0.95 & 0.92 & 0.91 & 0.81 & $-$ & $-$ & 0.85 & 0.80 & 0.96 & 0.93 & 0.97 & 0.96 \\
    \midrule
    \multirow{4}{*}{\begin{tabular}[l]{@{}c@{}}Delayed \\Messages \end{tabular}}
    & {0\%}
    & 0.91 & 0.82 & $-$ & $-$ & $-$ & $-$ & $-$ & $-$ & 0.80 & 0.55  & $-$ & $-$ & $-$ & $-$ & $-$ & $-$ \\

    & {1\%}
    & $-$ & $-$ & 0.90 & 0.81 & 0.90 & 0.81 & 0.90 & 0.81 & $-$ & $-$ & 0.79 & 0.53 & 0.80 & 0.54 & 0.79 & 0.54 \\

    & {5\%}
    & $-$ & $-$ & 0.90 & 0.80 & 0.90 & 0.80 & 0.90 & 0.81 & $-$ & $-$ & 0.78 & 0.51 & 0.80 & 0.54 & 0.79 & 0.53 \\

    & {10\%}
    & $-$ & $-$ & 0.89 & 0.79 & 0.89 & 0.80 & 0.90 & 0.81 & $-$ & $-$ & 0.77 & 0.47 & 0.80 & 0.54 & 0.78 & 0.53 \\

    & {20\%}
    & $-$ & $-$ & 0.89 & 0.78 & 0.90 & 0.80 & 0.90 & 0.81 & $-$ & $-$ & 0.74 & 0.41 & 0.80 & 0.51 & 0.76  & 0.51 \\
    
    \bottomrule
\end{tabular}
\vspace{-0.3cm}
\end{table*}

\subsection{Missing Patterns in Natural Occlusions}
In our experimental setup, natural occlusions are introduced in a principled manner by applying various missingness patterns.
Specifically, starting from complete BSM records in VeReMi, we employ standard statistical amputation techniques~\cite{Schouten2018}, namely, Missing Completely at Random (MCAR), Missing at Random (MAR), and Missing Not at Random (MNAR), to simulate the stochastic and structured nature of naturally occurring occlusions.  

In \textbf{MCAR}, missing values are injected uniformly at random across all features and samples, independently of both observed and unobserved data, producing missingness with no systematic pattern. To simulate \textbf{MAR}, the missingness probability is determined by the most correlated observed features via a logistic function across multiple data subsets. Under \textbf{MNAR}, features are randomly selected to induce missingness via a logistic function across multiple data subsets, simulating occlusions where a feature's value determines the likelihood of its own occlusion in the context of the VeReMi misbehavior types. For MAR and MNAR, we adapt the subset-based missingness generation approach of~\cite{9608954} to the specific feature structure of the VeReMi dataset. Moreover, missingness is induced across the complete dataset at missing ratios of 1\%, 5\%, 10\%, and 20\% to mimic realistic incomplete V2X environments.

\section{Performance Evaluation} \label{sec:res}

In this section, we evaluate the performance of the proposed DRL-based misbehavior detection framework across the three scenarios introduced in Section~\ref{sec:two}, using the VeReMi dataset. These scenarios enable a progressive assessment of the framework, ranging from naturally incomplete observations (SC1), to adversarial exploitation of environmental blind spots (SC2), and finally to deliberate manipulation of observability (SC3).
Results are benchmarked against a baseline classifier under identical missing data conditions. 

\subsection{Detection Performance}
Detection performance is evaluated in terms of \textit{Accuracy} (Acc) and \textit{F-score} (F1):
\setlength{\abovedisplayskip}{7pt}
\setlength{\belowdisplayskip}{7pt}
\begin{align}
\text{Acc} & = \frac{TP+TN}{TP+TN+FP+FN},
\label{eq:1} \enskip
\text{F1}  = 2\frac{RP}{R + P},
\end{align}
\noindent where \text{Acc} measures the ratio of correct predictions over all samples, and \text{F1} provides the harmonic mean of \textit{Precision} (P) and \textit{Recall} (R), making it the primary indicator of detection performance where both FP and FN rates are critical.

\subsection{Benchmark Scheme}
We select XGBoost \cite{xgboost16kdd} as our baseline classifier, a highly optimized Gradient Boosted Decision Tree (GBDT) framework that natively handles missing values. To ensure a fair comparison with our DRL-based approach, we employ a sliding-window mechanism similar to that used in the DRL setting.
Specifically, each temporal window is flattened into a single high-dimensional feature vector, and labels are assigned based on the final timestep within the window to maintain step-wise classification consistency with DRL. This setup provides XGBoost with identical historical context and input information.
We configure XGBoost's decision boundary hyperparameter to prioritize recall (minimizing missed attacks) while tolerating acceptable FPs, mirroring the DRL reward structure. This window-based strategy evaluates XGBoost’s ability to capture spatio-temporal dependencies within a sequential decision-making framework \cite{dietterich2002machine}. Such a design is particularly relevant for edge-based misbehavior detection\footnote{The computational overhead and real-time deployment feasibility of the proposed DRL agent for edge-based misbehavior detection were empirically assessed in our previous work \cite{sedar2024knowledge, asensio2024zsm}.},
where mission-critical applications must operate under limited data constraints. By flattening temporal sequences into high-dimensional spatial features, we assess whether the task can be effectively addressed through cross-feature correlations or whether the recurrent dynamics captured by DRL are necessary 
to effectively reason over
stealthy and persistent occlusions.


\subsection{Partial Observability under Natural Occlusions} 

To provide a comprehensive assessment under SC1, we evaluate the DRL-based framework's detection capability across multiple misbehaviors and missing patterns, as summarized in Table~\ref{tab:detection_perf_sc1}. The results consistently demonstrate the superior robustness of the DRL framework 
compared to XGBoost across all misbehavior types and missingness mechanisms. Under MCAR, DRL maintains strong performance even at 20\% missingness, achieving an accuracy of 0.82 and F1 of 0.58 for Constant Position, an accuracy of 0.89 and F1 of 0.81 for Constant Position Offset, an accuracy of 0.92 and F1 of 0.86 for Random Position, an accuracy of 0.89 and F1 of 0.77 for Random Speed, and an accuracy of 0.89 and F1 of 0.78 for Delayed Messages. In contrast, XGBoost exhibits significant performance degradation at 20\% MCAR, with F1 dropping to the range of 0.41-0.80 under the same conditions.

In the presence of MNAR, where missingness depends on the feature values themselves, DRL achieves consistently high accuracy (0.90-0.96) and F1 scores (0.80-0.93) across all misbehavior types at 20\% missingness. XGBoost shows markedly lower detection performance, particularly for Constant Position Offset (an F1 of 0.70) and Delayed Messages (an F1 of 0.51). The MAR mechanism reveals similar trends, with DRL maintaining robust detection capabilities with F1 scores in the range of 0.81-0.93 at a 20\% missing ratio, while XGBoost drops to F1 scores between 0.51 and 0.79 for most misbehavior types, except for Random Speed, where it retains an F1 of 0.96.    

\begin{table*}[!ht]
\centering
\caption{Comparative performance of DRL and XGBoost under evasion via natural occlusions in SC2.}
\label{tab:drl_xbg_evasion}
\small
\resizebox{1.7\columnwidth}{!}{%
\begin{tabular}{l c |ccc |ccc}
    \toprule
    \rowcolor{rowgray}
    & & \multicolumn{3}{c|}{\textbf{DRL}} & \multicolumn{3}{c}{\textbf{XGBoost}} \\ 
    \cmidrule(lr){3-5} \cmidrule(lr){6-8}
    \rowcolor{rowgray}
     & \textbf{Missingness} 
    & \textbf{Acc} & \textbf{F1} & \textbf{ASR (\%)} 
    & \textbf{Acc} & \textbf{F1} & \textbf{ASR (\%)} \\
    \midrule
    \textbf{No Attack} & 0\% & 1.00 & 1.00 & 0.00 & 0.99 & 0.98 & 0.00 \\
    \midrule
    
    \rowcolor{rowgray} 
    \textbf{Evasion+MCAR} & & & & & & & \\
    & 1\%  & 0.90 $\pm$ 0.02 & 0.83 $\pm$ 0.03 & 71.57 $\pm$ 1.49 & 0.98 $\pm$ 0.01 & 0.97 $\pm$ 0.02 & 4.78 $\pm$ 0.33 \\
    & 5\%  & 0.92 $\pm$ 0.01 & 0.86 $\pm$ 0.02 & 65.19 $\pm$ 1.58 & 0.95 $\pm$ 0.01 & 0.92 $\pm$ 0.02 & 10.62 $\pm$ 0.29 \\
    & 10\% & 0.92 $\pm$ 0.01 & 0.87 $\pm$ 0.02 & 62.05 $\pm$ 1.57 & 0.92 $\pm$ 0.01 & 0.86 $\pm$ 0.02 & 18.47 $\pm$ 0.23 \\
    & 20\% & 0.93 $\pm$ 0.01 & 0.88 $\pm$ 0.01 & 58.48 $\pm$ 0.77 & 0.86 $\pm$ 0.01 & 0.74 $\pm$ 0.02 & 34.59 $\pm$ 0.55 \\

    \midrule
    \rowcolor{rowgray} 
    \textbf{Evasion+MNAR} & & & & & &  &\\
    & 1\%  & 0.89 $\pm$ 0.02 & 0.82 $\pm$ 0.04 & 73.25 $\pm$ 1.86 & 0.99 $\pm$ 0.01 & 0.98 $\pm$ 0.02 & 4.14 $\pm$ 5.07 \\
    & 5\%  & 0.89 $\pm$ 0.02 & 0.80 $\pm$ 0.04 & 75.44 $\pm$ 1.67 & 0.97 $\pm$ 0.01 & 0.95 $\pm$ 0.01 & 16.43 $\pm$ 1.29 \\
    & 10\% & 0.90 $\pm$ 0.02 & 0.82 $\pm$ 0.04 & 72.82 $\pm$ 3.93 & 0.96 $\pm$ 0.02 & 0.93 $\pm$ 0.03 & 15.35 $\pm$ 8.81 \\
    & 20\% & 0.90 $\pm$ 0.02 & 0.83 $\pm$ 0.04 & 67.33 $\pm$ 7.99 & 0.96 $\pm$ 0.03 & 0.92 $\pm$ 0.05 & 8.03 $\pm$ 12.95 \\
    
    \midrule
    \rowcolor{rowgray} 
    \textbf{Evasion+MAR} & & & & & & &\\
    & 1\%  & 0.89 $\pm$ 0.02 & 0.81 $\pm$ 0.04 & 73.63 $\pm$ 1.69 & 0.97 $\pm$ 0.01 & 0.96 $\pm$ 0.02 & 0.27 $\pm$ 0.35 \\
    & 5\%  & 0.89 $\pm$ 0.02 & 0.81 $\pm$ 0.03 & 74.29 $\pm$ 1.10 & 0.97 $\pm$ 0.01 & 0.95 $\pm$ 0.02 & 0.37 $\pm$ 0.13 \\
    & 10\% & 0.89 $\pm$ 0.02 & 0.81 $\pm$ 0.03 & 75.08 $\pm$ 0.99 & 0.96 $\pm$ 0.02 & 0.94 $\pm$ 0.03 & 0.30 $\pm$ 0.05 \\
    & 20\% & 0.88 $\pm$ 0.01 & 0.78 $\pm$ 0.03 & 77.73 $\pm$ 1.02 & 0.95 $\pm$ 0.02 & 0.92 $\pm$ 0.03 & 0.38 $\pm$ 0.09 \\
    \bottomrule
\end{tabular}
}
\vspace{-0.3cm}
\end{table*}

Notably, DRL exhibits varying sensitivity to missingness proportions across misbehavior types (Table~\ref{tab:detection_perf_sc1}), with F1 degradation ranging from 5\% to 41\% when increasing from 0\% to 20\% missingness under the most challenging pattern in MCAR. The detection of Delayed Messages shows minimal degradation with F1 decreasing from 0.82 to 0.78 (5\% drop), while the detection of Constant Position misbehavior shows the highest performance degradation with F1 decreasing from 0.98 to 0.58 (41\% drop). For Random Position and Random Speed, DRL demonstrates moderate robustness with 11\% and 23\% degradation, respectively. XGBoost suffers comparable or greater degradation under the same conditions. Constant Position detection declines from F1 0.88 to 0.59 (33\% drop), Random Position from 0.78 to 0.60 (23\% drop), and Delayed Messages from 0.55 to 0.41 (25\% drop).

DRL demonstrates significantly stronger robustness under structured missingness patterns. Under MAR, F1 scores range from 0.81 to 0.93 at 20\% missingness (1.2\%-19\% drop), while under MNAR, F1 ranges from 0.80 to 0.93 (2.4\%-9.3\% drop). Notably, MNAR exhibits the lowest worst-case performance degradation for the DRL framework across all misbehavior types shown in Table~\ref{tab:detection_perf_sc1}. In comparison, XGBoost shows F1 scores ranging from 0.51 to 0.96 under MAR (0\%-10\% drop) and from 0.51 to 0.93 under MNAR (3\%-8\% drop). This adaptive robustness stems from DRL's ability to capture spatio-temporal dependencies within a sequential decision-making framework, which effectively exploits the structure in MAR and MNAR patterns to mitigate the impact of incomplete observations. In contrast, MCAR's completely random missingness disrupts temporal continuity severely, posing greater challenges even for sequential models.

\subsection{Partial Observability under Evasion Attacks} 
We evaluate the robustness of DRL under evasion attacks (SC2) considering the Random Speed misbehavior, where both DRL and XGBoost obtained comparable performance across different missingness proportions under natural occlusions in SC1 (Table~\ref{tab:detection_perf_sc1}). Their adversarial robustness is summarized in Table~\ref{tab:drl_xbg_evasion}. As shown in the table, DRL exhibits significant vulnerability with attack success rates (ASR) consistently exceeding 70\% even at minimal missingness rates of 1\% under all three evasions via natural missingness patterns. At the highest missingness of 20\%, DRL's ASR ranges from 58.48\% (MCAR) to 77.73\% (MAR), demonstrating that adversaries successfully evade detection in approximately 75\% of attempts by exploiting natural occlusions. Despite maintaining relatively high accuracy between 0.88 and 0.93, this vulnerability arises from DRL's reliance on sequential temporal dependencies, which makes it challenging to distinguish between legitimate incomplete observations and adversarial manipulations that exploit natural occlusions as a camouflage.   

\begin{table}[!t]
\centering
\caption{Performance comparison under feature suppression attack (SC3).}
\label{tab:drl_xgb_supp}
\small
\resizebox{\columnwidth}{!}{%
\begin{tabular}{lcccc}
\toprule
 & \textbf{Intensity~($\lambda$)} & \textbf{Acc} & \textbf{F1} & \textbf{F1 Drop (\%)} \\
\midrule
\rowcolor{rowgray} 
\multicolumn{5}{l}{\textbf{DRL}} \\
\quad No Attack & 0.0 & 1.00 & 1.00 & - \\
\quad Weak & 0.1 & 0.99 $\pm$ 0.00 & 0.98 $\pm$ 0.01 & -2.1 \\
\quad Medium & 0.2 & 0.98 $\pm$ 0.00 & 0.97 $\pm$ 0.01 & -2.8 \\
\quad Strong & 0.3 & 0.98 $\pm$ 0.00 & 0.96 $\pm$ 0.01 & -3.9 \\
\quad Stronger & 0.4 & 0.97 $\pm$ 0.00 & 0.95 $\pm$ 0.01 & -5.1 \\
\quad Strongest & 0.5 & 0.96 $\pm$ 0.04 & 0.93 $\pm$ 0.07 & -7.3 \\
\midrule
\rowcolor{rowgray} 
\multicolumn{5}{l}{\textbf{XGBoost}} \\
\quad No Attack & 0.0 & 0.99 & 0.98 & - \\
\quad Weak & 0.1 & 0.99 $\pm$ 0.01 & 0.98 $\pm$ 0.02 & +0.1 \\
\quad Medium & 0.2 & 0.88 $\pm$ 0.03 & 0.83 $\pm$ 0.04 & -15.8 \\
\quad Saturated & 0.3 & 0.88 $\pm$ 0.03 & 0.83 $\pm$ 0.04 & -15.7 \\
\quad Saturated & 0.4 & 0.88 $\pm$ 0.03 & 0.83 $\pm$ 0.04 & -15.6 \\
\quad Saturated & 0.5 & 0.88 $\pm$ 0.03 & 0.83 $\pm$ 0.04 & -15.5 \\
\bottomrule
\end{tabular}
}
\end{table}

In contrast, XGBoost shows notably higher robustness across all evasion attacks via natural missingness (Table~\ref{tab:drl_xbg_evasion}). Under evasion via MAR, XGBoost achieves ASR below 0.4\% across all missingness proportions. Against challenging MCAR attacks at 20\% missingness, XGBoost limits ASR to 34.59\%, an approximately twofold improvement over DRL. In particular, the robustness gap is substantial for evasion via MAR patterns, where XGBoost's 0.38\% of ASR versus DRL's 77.73\% yields a 205$\times$ advantage in attack resistance. This robustness arises from XGBoost's tree ensemble processing temporal windows as flattened high-dimensional feature vectors, treating each window independently rather than enforcing sequential dependencies that make DRL vulnerable. XGBoost's ensemble architecture dilutes the impact of adversarial perturbations across the larger feature space and shows greater resistance to distribution-shift attacks compared to the DRL's sequential approach. 

\subsection{Partial Observability under Feature Suppression}
\begin{figure}[t!]
    \centering
    \begin{subfigure}[b]{0.242\textwidth}
        \centering
        \includegraphics[width=\textwidth]{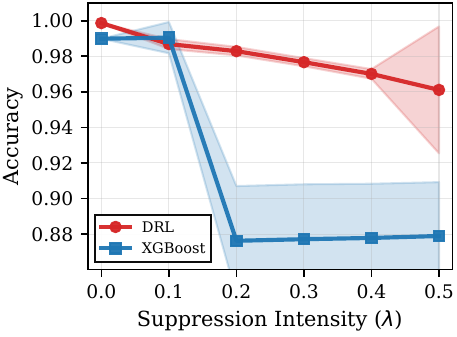}
    \end{subfigure}\hfill
    \begin{subfigure}[b]{0.242\textwidth}
        \centering
        \includegraphics[width=\textwidth]{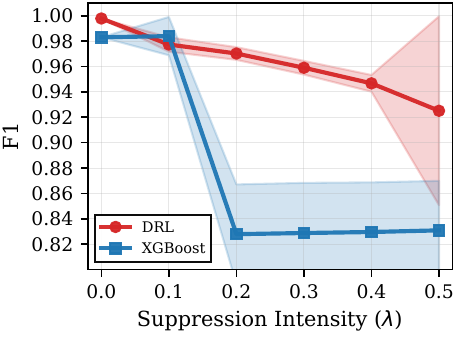}
    \end{subfigure}

 \caption{Performance degradation under feature suppression attack (SC3).}
    \label{fig:drl_xgb_perf_supp}
    \vspace{-0.35cm}
\end{figure}

To assess the robustness of DRL under adversarial feature suppression (SC3), we again consider the Random Speed misbehavior, where both classifiers obtained comparable performance in SC1 (Table~\ref{tab:detection_perf_sc1}), and compare their adversarial robustness. Feature suppression attacks on complete data (i.e., without natural occlusions) reveal contrasting vulnerability patterns between DRL and XGBoost, as shown in Table~\ref{tab:drl_xgb_supp} and Fig.~\ref{fig:drl_xgb_perf_supp}. DRL exhibits gradual performance degradation as suppression intensity ($\lambda$) increases from 0.1 to 0.5, with accuracy declining from 0.99 to 0.96 and F1 score from 0.98 to 0.93 (7.3\% drop). As illustrated in Fig.~\ref{fig:drl_xgb_perf_supp}, this gradual degradation follows a near-linear decline, reflecting how artificial missingness through targeted feature suppression progressively disrupts the spatio-temporal patterns captured by DRL's recurrent dynamics, which are essential for coherent state representation across sequential timesteps. The DRL agent's ability to accumulate evidence across sequential timesteps weakens as suppression intensity increases, leading to increasing FNs and a gradual decline in F1 performance.

In contrast, XGBoost exhibits a significant performance drop between intensities 0.1 and 0.2, clearly visible in both accuracy and F1 plots (Fig.~\ref{fig:drl_xgb_perf_supp}). At intensity 0.1, XGBoost maintains near-baseline performance with accuracy of 0.99 and F1 of 0.98, but at intensity 0.2, performance drops abruptly to accuracy of 0.88 and F1 of 0.83 (15.8\% degradation), then plateaus across higher intensities with minimal further degradation. This sudden drop occurs as XGBoost's tree ensemble loses critical discriminative features once suppression intensity reaches approximately 0.2. Unlike DRL's gradual degradation, XGBoost's F1 collapses from 0.98 to 0.83, exhibiting an overly conservative classification where the model flags many benign samples as malicious. These findings reveal that DRL's graceful degradation to F1 of 0.93 at intensity 0.5 offers more predictable behavior than XGBoost's abrupt F1 collapse to 0.83, making it preferable for mission-critical V2X deployments. Despite DRL's more favorable degradation pattern, enhanced defensive mechanisms are required to strengthen its robustness against targeted feature suppression attacks.

\section{Conclusions} \label{sec:concl}
This paper investigated the problem of misbehavior detection with partially observable V2X measurements, a setting often overlooked in existing ML-based detection methods.
This partial observability, owing to sensor imperfections, intermittent connectivity, and/or environmental occlusions, can be further leveraged by malicious actors via deliberate exploitation of naturally missing regions and adversarial feature suppression.
Departing from conventional supervised classification methods, we proposed a DRL-based framework capable of learning robust detection policies under incomplete observations.

Experimental results on the VeReMi dataset reveal three key insights. First, under natural occlusions, the proposed DRL approach consistently outperforms a powerful XGBoost baseline across MCAR, MAR, and MNAR patterns, demonstrating the benefit of sequential policy learning in structured partially observable environments. Second, under evasion attacks that exploit natural missingness, DRL exhibits significant vulnerability, with high ASRs despite maintaining high accuracy. This shows that adversaries can leverage natural occlusions as a camouflage mechanism against sequential detectors. Third, under adversarial feature suppression, DRL shows gradual degradation in decision-making, whereas XGBoost exhibits abrupt performance collapse once critical features are suppressed.
%
Future work will explore the use of adversarial training to harden the DRL agent's sequential policy against the camouflage effects of natural occlusions.

\bibliographystyle{IEEEtran}
\bibliography{refs}

\end{document}